\documentstyle[aps,prl]{revtex}
\preprint{}
\begin{document}
\title{Imaginary Potential as a Counter of Delay Time for Wave Reflection 
from a 1D Random Potential}
\author{S.Anantha Ramakrishna\footnote{ E-mail : sar@rri.ernet.in} 
and N.Kumar\footnote{E-mail : nkumar@rri.ernet.in}\footnote{Work supported partially by the 
Max-Planck Institut f\"{u}r Komplexer Systeme, 01187 Dresden, Germany}}
\address{Raman Research Institute, C.V.Raman Avenue, Bangalore - 560 080, India}
\date{\today}
\maketitle
\begin{abstract}
We show that the delay time distribution for wave reflection from a 
one-dimensional (1-channel) random potential is related directly to that of the 
reflection coefficient, derived  with an arbitrarily small but uniform 
imaginary part added to the random potential. Physically, the reflection 
coefficient, being exponential in the time dwelt in the presence of the 
imaginary part, provides a natural counter for it. The delay time 
distribution then follows straightforwardly from our earlier  results for 
the reflection coefficient, and coincides  with the distribution 
obtained recently by Texier and Comtet [C.Texier and A. Comtet,
Phys.Rev.Lett. {\bf 82}, 4220 (1999)],
with all moments infinite. Delay time distribution  for a random amplifying
medium is then derived . In this case, however, all moments work  out to be
finite.\\
\end{abstract}
%\newpage
        When a wavepacket centered at an energy $E$ is scattered elastically 
from a scattering potential, it suffers a time delay before spreading out  
dispersively. This delay is related to the time for which the wave dwells in 
the interaction region. For the general case of a scatterer coupled to N open 
channels leading to the continuum, one defines the phase-shift time delays through 
the Hermitian energy derivative of the $S$-matrix, $-i \hbar S^{-1} 
\partial S / \partial E $, whose eigenvalues give the proper delay times.
These delay times then averaged over the N-channels give the Wigner-Smith delay 
times introduced first by Wigner\cite{wigner} for the 1-channel case, and 
generalized later by Smith\cite{smith} to the case of N open channels. 
Thus scattering delay time is the single most important quantity describing 
the time-dependent aspect {\it i.e.,} physically, the reactive aspect of the 
scattering in open quantum systems, {\it e.g.} the chaotic microwave cavity 
and the quantum billiard (whose classical motion is chaotic) and the solid-state 
mesoscopic dots coupled capacitively to open leads terminated in the reservoir. 
The delay time is however not self averaging and one must have its full probability
distribution over a statistical ensemble of random samples. The latter may 
be related ergodically  to the ensembles generated parametrically {\it e.g.} 
by energy $E$ variation over a sufficient interval. Thus we have the random 
matrix theory (RMT) for circular ensembles of the $S$-matrix giving delay 
times for all the three Dyson Universality classes for the case of a 
chaotic cavity connected to a single open channel \cite{gopar}. 
Generalization to the case of N channels corresponded to the Laguarre 
ensemble \cite{brouwer} of RMT. The RMT approach has been treated earlier 
through the supersymmetric  technique for the case of a quantum chaotic 
cavity having a few equivalent open channels \cite{fyodorov1}. However it 
has been suspected for quite sometime that the RMT based results and 
the universality claimed thereby may not extend to a strictly 
1-dimensional random system where Anderson localization dominates, and 
that the 1D random system may constitute  after all a different universality 
class \cite{fyodorov2}. 
This important problem has been re-examined recently by Texier 
and Comtet \cite{texier} who have derived the delay time distribution 
for a 1D conductor with the Frish-Lloyd  model randomness in the limit 
of high energy / weak disorder and the sample length $>>$ the localization 
length. The universality of the distribution is amply supported by numerical
simulations for different models of disorder \cite{texier,joshi}. \\

In this work we re-examine this question of universality of the delay time
distribution for a 1D random system and relate it to the universality of 
distribution of the reflection coefficient, a quantity that we have direct
access to from our earlier work \cite{pradhan}. To this end we introduce a 
novel counter that literally clocks the time dwelt by the wave in the 
scattering region, obviating the need for calculating the energy derivative 
of the phase shift \cite{jayanna}. This involves adding formally an arbitrarily small 
but uniform imaginary part $iV_{i}$ to the 1-D random potential $V_{r}$. Now, 
the reflection coefficient, being exponential in the time dwelt  in the 
scattering region in the presence of $iV_{i}$, provides a literal `counter' 
for this time.The distribution derived by us agrees exactly with the universal 
time-delay distribution of Texier and Comtet \cite{texier}. Besides, our 
new technique allows us to treat the time-delay distribution for the 
important case of light  
reflected from a random amplifying medium equally well. In this case however, 
unlike the case for the passive random medium, all moments of the delay time 
are finite for long samples. \\  

Consider first the electronic case for a 1D disordered sample of length $L$ 
having a random potential $V_{r}~$, $~0 \le x \le L$, and connected to 
infinitely long perfect leads at the two ends. Let the electron wave of 
energy $E = \hbar^{2} k^{2} /2 m$ be incident from the right at $x = L$, 
and be partially reflected with a complex amplitude reflection coefficient 
$R(L) = \vert R(L) \vert \exp\left( i\theta (L) \right) $ and 
$\vert R(L) \vert^{2} = r(L)$, the real reflection coefficient. Inside the 
sample we have the Schr\"{o}dinger equation,
\begin{equation}
\label{wave}
\frac{d^{2} \psi (x)}{d x^{2} } + k^{2} \left( 1 + \eta_{r}(x) \right)
\psi (x) = 0 ,
\end{equation}
with  $\eta_{r} (x) = - V_{r} (x) / E $. \\

As we will be interested in the reflection coefficient, it is apt to follow 
the invariant imbedding technique \cite{pradhan,jayanna,heinrichs,rammal}
and reduce the Schr\"{o}dinger equation(\ref{wave}) to an equation for the emergent 
quantity $R(L)$ :
\begin{equation}
\label{reflect}
\frac{dR(L)}{dL} = 2 i k R(L) + \frac{ik}{2} \eta_{r}(L) \left( 1+ R(L) \right)^{2} .
\end{equation}
We now introduce a uniform imaginary part $iV_{i}$, with $V_{i} > 0$, and 
accordingly define  $\eta (L) = \eta_{r} + i \eta_{i}$, with 
$\eta_{i}=-V_{i}/E$. For analytical treatment, we take for $V_r(x)$ a 
gaussian delta-correlated random potential (the Halperin model) with 
$ < \eta_{r}(L) > = 0$ and  $< \eta_{r}(L) \eta_{r}(L') > = \Delta^{2} 
\delta(L-L')$. The Fokker-Planck equation corresponding to the stochastic 
equation(\ref{reflect}) can be solved analytically in the limit 
$L \rightarrow \infty$ giving \cite{pradhan}: 
\begin{eqnarray}
\label{pofr}
P_{\infty} (r) & = & \frac{ D \exp \left( - \frac{D}{r-1} \right) }{(r-1)^{2} }
~~~~, ~~~~~~~~~~~~~~~ r \ge 1   \\
       & = &  0 ~~~~~~~~~~~~~~~~~~~~~~, ~~~~~~~~~~~~~~~~ r < 1 \nonumber
\end{eqnarray}
with $D = (4 V_{i})/(E \Delta^{2} k)$. This result is obtained in the high 
energy / weak disorder limit. Now, clearly for a passive medium, {\it i.e.,}
with $V_{i} = 0$, the distribution $P_{\infty}(r)$ must collapse to a 
delta-function $\delta(r-1)$ as $L \rightarrow \infty$. However, with $V_{i}
\ne 0 $, for a short dwell time $T$ in the sample, the reflection coefficient
$r = \vert R \vert^{2} = \exp(2 V_{i}T / \hbar)$, giving $r-1=2V_{i}T/\hbar$
to first order in $V_{i}$ as $V_{i}$ is taken to be arbitrarily small. Thus,
$P_{\infty}(r)$ can at once be translated into the dwell time distribution 
$P^{0}_{\infty} (\tau )$:
\begin{equation}
\label{poft}
P^{0}_{\infty}( \tau ) =  \frac{\alpha}{\tau^{2}} \exp \left(- \frac{\alpha}
{\tau} \right) ,
\end{equation}
where $\alpha = 2 (\Delta^{2} k)^{-1}$ and the dimensionless time $\tau = 
E T / \hbar $. This is precisely the result of Texier and Comtet \cite{texier}.
Note that $V_{i}$, the counter, drops out in the limit $V_{i} \rightarrow 0$,
as it should. It should also be noted that the invariant imbedding equation 
for the energy derivative of the phase shift\cite{jayanna} also yields the 
same result for the delay time distibution when the high energy limit ( $
k \rightarrow \infty$ while keeping $V_{r}/k$ constant) is explicitly taken. 
This again re-confirms our delay time distribution given above.\\

At this point, it is perhaps apt to demystify our time delay counter {\it viz.}
the introduction of an imaginary potential ($V_{i}$) in the limit $V_{i} 
\rightarrow 0$ as a mathematical artifice for the electronic case, in terms of the well-known analytic
property of the S-matrix, corresponding to wave reflection from the 1-D 
infinitely long disordered system. The S-matrix in this case is simply the 
complex \underline{amplitude} reflection coefficient, $R(E) = \exp [i\theta (E)]$
with $\vert R \vert^{2} = 1$ for real $E$. Now from the analyticity of the S-matrix 
in the complex energy plane, we have $\partial (Re~\theta)/\partial (Re~E) = 
\partial (Im~\theta)/\partial (Im~E)$, where $Re$ and $Im$ denote the real and 
the imaginary parts respectively. As we approach the real axis, {\it i.e.,} in
the limit $Im E \rightarrow 0$, we have $\partial (Re~\theta)/\partial (Re~E) =
T/\hbar$ (Wigner time delay), while $\partial (Im~\theta)/\partial (Im~E)
\rightarrow Im~\theta/V_{i}$ as $V_{i} \rightarrow 0$ (along with $Im~\theta$).
Thus we have $\vert R \vert^{2} = \exp [2 V_{i} T/\hbar]$ giving $\vert R \vert^{2} 
- 1 = 2 V_{i} T/\hbar$ in the limit $V_{i} \rightarrow 0$ (the latter corresponds to 
treating our electronic problem as a limit of vanishing imaginary
part of the scattering potential). This is what has been used above to obtain the
delay time distribution from  the reflection coefficient distribution given by
eqn.(\ref{pofr}) in the limit  $V_{i} \rightarrow 0$.\\

Encouraged by this result for for the electronic case, we now turn to 
the case of a light wave reflected from a Random Amplifying Medium. The latter 
is recieving much attention in recent years in the context of random lasers 
\cite{lawandy,wiersma,hema}. To fix ideas, consider the case of a single mode 
optical fibre doped with $Er^{3+}$, say, optically pumped and intentionally  
disordered refractively. All we have to do now is to keep $V_{i}$ finite, 
a measure of medium gain, and use $T = (\hbar / 2)~~ (\partial \ln r / \partial
V_{i})$ for the dwell time, and translate $P_{\infty} (r)$ into $P_{\infty}
(\tau )$ :
\begin{equation}
\label{tdelamp}
P_{\infty} ( \tau ) = (D \xi) \frac{ \exp \left( -\frac{D}{e^{\xi \tau} 
 - 1} \right)}{(e^{\xi \tau} - 1)^{2}} e^{\xi \tau} ,
\end{equation}
where $\xi = 2 V_{i} /E$. 
Again, $P_{\infty}$ vanishes in the limit $\tau \rightarrow \infty$ as also
for $ \tau \rightarrow 0$. 
Also, $P_{\infty}(\tau) \rightarrow P^{0}_{\infty}$ as $V_{i} 
\rightarrow 0$. 
All moments $< \tau^{n} >$  are however finite in 
this case. An explicit expression can be obtained for the first moment as :
\begin{equation}
\langle \tau \rangle = \frac{1}{\xi} \left[ \ln D +C - e^{D} Ei(-D) \right] ,
\end{equation}
where $C$ is the Euler's constant \cite{gradshteyn} and $Ei$ is the exponential
integral \cite{gradshteyn}. This expression diverges as $V_i \rightarrow 0$. 
In  Fig.1, we show the delay time distributions given by eq.(\ref{tdelamp}) 
for different values of the parameter $\xi$ keeping $\alpha$ fixed 
corresponding to different values of of the imaginary potential $V_i$
while keeping the disorder fixed.  

%\begin{figure}
%\vspace{5mm}
%\epsfxsize=250pt
%\begin{center}{\mbox{\epsffile{tdelfig.ps}}}
%\end{center}
%\vspace{5mm}
%\caption{The delay time distribution from an amplifying medium}
%\vspace{0.5cm}
%%\hrule{iijjjj}
%\label{fig1}
%\end{figure}

Several interesting points are to be noted here. The counter introduced by us 
literally counts the dwell time in the interaction region  for 
total reflection in the 1D, {\it i.e.,}1-channel case. Large delay time is 
dominated by the dwell time when the wave penetrates deeper into the sample, 
which is true at high energy/low disorder. It is this `equilibrated' part of 
the reflected wave, and not the prompt part that is expected to give 
universality.  Hence the universal $1/ \tau^{2}$ tail in equation(\ref{poft}).
Indeed, the universality of the delay-time distribution directly reflects
that of the reflection coefficient given by equation(\ref{pofr}) 
\cite{zhang,beenakker,jiang}. Indeed, 
we have verified that equation(\ref{pofr}) is obtained for telegraph disorder
also. It is to be remarked here that this universal delay time distribution
as in equation(\ref{poft}), is not obtained for a chaotic cavity connected to a 
reservoir by a single open channel \cite{gopar}. Here the localization 
picture may not hold. As for the finiteness of all the moments $<\tau^{n} >$
for the case of the random anplifying medium, it is quite consistent with the 
known fact that the amplification enhances localization and thus prevents
deep penetration in the random sample. Of course, there is also an  
enhanced prompt part of reflection resulting 
from the increased refractive index mismatch in respect of its imaginary 
part at the sample-lead interface.\\

In conclusion, we have introduced a novel `counter' that measures the dwell 
time in the scattering medium. We have used it successfully to derive 
the delay time distribution in terms of that of the reflection time. Both 
passive and amplifying media have been treated. Our counter can be used
equally well in principle to calculate the traversal time for the problem of tunneling
across a potential barrier \cite{landauer,jayanna2} \\

\acknowledgements
One of us (N.K) acknowledges hospitality offered by the Max-Planck Institut
f\"{u}r Komplexer Systeme, Dresden 01187, Germany 
during the course of this work, May-June 1999.  We thank Y.V. Fyodorov 
for helpful comments. \\

\vspace{1cm}
\begin{center}
{\bf Figure Captions}
\end{center}

FIG. 1.  The delay time distribution from an amplifying medium.

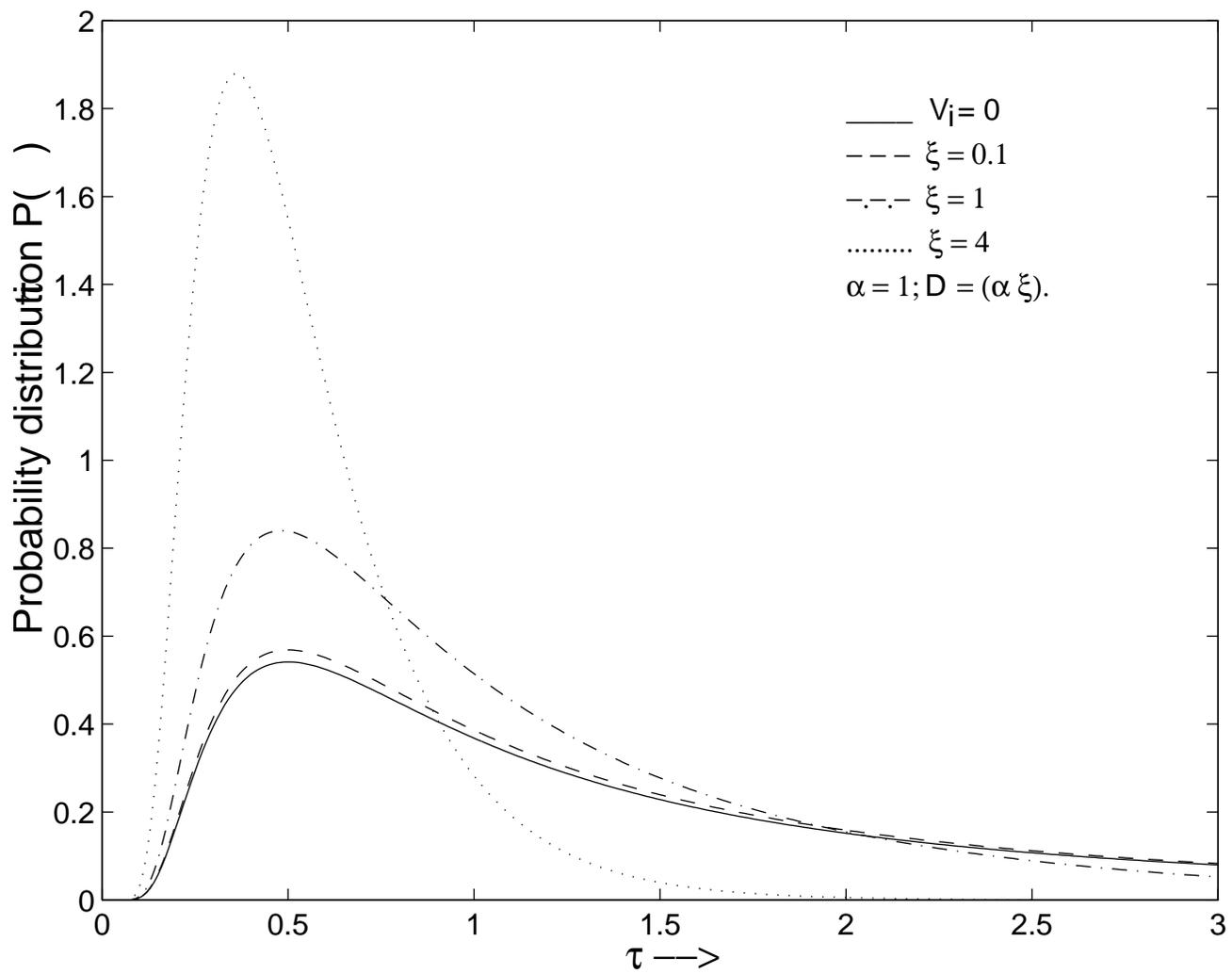
\end{document}